\input{aipcheck}

\documentclass[
    ,final            
  ]
  {aipproc}

\layoutstyle{6x9}

\begin{document}

\title{The formation of supermassive black holes in the first galaxies}


\author{Dominik~R.~G.~Schleicher}{
  address={ESO Garching, Karl-Schwarzschild-Str. 2, 85748 Garching bei M\"unchen, Germany (dschleic@eso.org); \\Leiden Observatory, Leiden University, P.O.Box 9513, NL-2300 RA Leiden, the Netherlands}
}

\author{Robi~Banerjee$^{\dagger}$, Sharanya~Sur$^{\dagger}$, Simon~C.~O.~Glover}{
  address={Zentrum f\"ur Astronomie der Universit\"at Heidelberg, Institut f\"ur Theoretische Astrophysik, Albert-Ueberle-Str.~2, D-69120 Heidelberg, Germany}
  }
  \author{Marco~Spaans}{
  address={Kapteyn Astronomical Institute, University of Groningen, P.O. Box 800, 9700 AV, Groningen, the Netherlands}
}
  
%

  \author{Ralf S. Klessen$^{\dagger}$}{
  address={
  Kavli Institute for Particle Astrophysics and Cosmology, Stanford University, Menlo Park, CA 94025, U.S.A.}
  }

\begin{abstract}
We discuss the formation of supermassive black holes in the early universe, and how to probe their subsequent evolution with the upcoming mm/sub-mm telescope ALMA. We first focus on the chemical and radiative conditions for black hole formation, in particular considering radiation trapping and molecular dissociation effects. We then turn our attention towards the magnetic properties in the halos where the first black holes form, and show that the presence of turbulence may lead to a magnetic dynamo, which could support the black hole formation process by providing an efficient means of transporting the angular momentum. We finally focus on observable properties of high-redshift black holes with respect to ALMA, and discuss how to distinguish between chemistry driven by the starburst and chemistry driven by X-rays from the black hole.
\end{abstract}

\maketitle


\section{Chemical and radiative conditions}
The presence of supermassive black holes at $z>6$ raises the question how they formed so quickly. Forming them from stellar progenitors seems difficult, as the first stars were thought to be very massive, thus giving rise to strong radiation feedback \citep{Abel02, Bromm04}. The dark matter halos in which they form are thus photo-evaporated at the end of their lifetime, leaving little gas available for subsequent accretion \citep{Johnson07, Milosavljevic08, Alvarez09}.

As an alternative, it was suggested that more massive seeds could form in the first atomic cooling halos \citep[e.g.][]{Eisenstein95, Bromm03, Begelman06, Spaans06}. These systems have masses $M > 5 \times 10^{7} [(1 + z) / 10]^{-3/2} \: {\rm M_{\odot}}$ and thus virial temperatures $T_{\rm vir} > 10^{4} \: {\rm K}$. Virialization shocks may thus sufficiently heat the gas such that cooling with the atomic hydrogen lines is relevant. To form a massive central clump, fragmentation should be suppressed. This is possible for instance if cooling is inefficient, leading to a stiff equation of state \citep{LiMacLowKlessen05}. The large hydrogen column densities may lead to efficient trapping of Lyman-$\alpha$ photons and effectively shut down Lyman-$\alpha$ cooling \citep{Spaans06}, while photodissociation may  suppress cooling via molecular hydrogen \citep[e.g.][]{Bromm03,Regan09, Shang09}. We investigate these effects in the presence of additional cooling mechanisms expected to be present in primordial gas, in particular chemical cooling channels like H$^-$ formation cooling and H$_2$ collisional dissociation cooling \citep{Omukai01, Glover09}. 

\begin{figure}
  \includegraphics[scale=0.45]{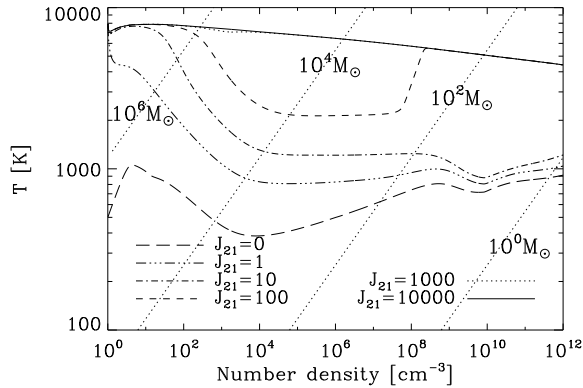}
  \caption{Temperature evolution as a function of density for different values of $J_{21}$. The thin dotted lines indicate
lines of constant Jeans mass. }
\label{fig:temp}
\end{figure}

For the photodissociating background, we adopt a blackbody spectrum with temperature $T=10^5$~K, which we normalize at $13.6$~eV with the parameter $J_{21}$, where $J_{21}=1$ corresponds to a UV background strength of $10^{-21}$~erg~s$^{-1}$~cm$^{-2}$~sr$^{-1}$~Hz$^{-1}$. The detailed approach is described by \citet{Schleicher10b}. Work by \citet{Dijkstra08} showed that even $J_{21}>1000$ may occur frequently enough to explain the abundance of supermassive black holes at $z\sim6$. We explore values up to $J_{21}=10^4$ and show the resulting temperature evolution in Fig.~\ref{fig:temp}, which is independent of the adopted hydrogen column density. Even for the extreme cases, the temperature keeps decreasing with density, so that fragmentation cannot be excluded on thermodynamical grounds alone.

\section{Magnetic conditions and dynamo effects}
\begin{figure}
  \includegraphics[scale=0.45]{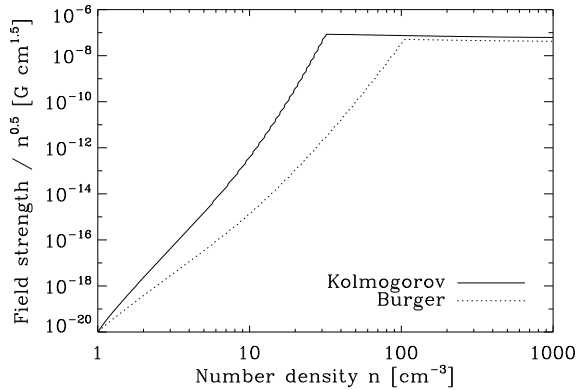}
  \caption{Magnetic field amplification in the early collapse phase, as a function of the mean density in the protostellar core. }
\label{fig:bfield}
\end{figure}

An additional mechanism that may stabilize the gas against fragmentation and enhance the transport of angular momentum are magnetic fields. As long as they are below equipartition, they can be well-described with the ideal MHD equations due to the high ionisation degree of the gas \citep{Maki04, Glover09}. Numerical studies showed that the gas during the formation of the first stars is turbulent \citep{Abel02, Yoshida08}, which can be seen in the inhomogeneous density distribution in the protostellar core and from the specific angular momentum profiles, which are sub-Keplerian because of the turbulent angular momentum transport. Under such conditions, a small-scale dynamo may operate \citep[e.g.][]{Kazantsev68, Brandenburg05}. It has been suggested to be operational in the first galaxies \citep{Arshakian09}, and the corresponding field amplification during the formation of the first stars in minihalos and atomic cooling halos has been explored in a semi-analytic framework by \citet{Schleicher10c}. In this framework, we model the growth of the magnetic field on different scales for a Kolmogorov or a Burgers type spectrum. As predicted for the small-scale dynamo, we assume that on each scale, it is amplified by the corresponding eddy-turnover scale. The smallest scales therefore grow fastest and saturate first, but even scales corresponding to $1/10$ of a Jeans length will eventually saturate, as the eddy-turnover time is significantly smaller than the free-fall time. The evolution of the magnetic field strength obtained with this model is shown in Fig.~\ref{fig:bfield}. We are currently probing this scenario with numerical MHD simulations following the collapse of primordial clouds. Additional implications due to the magneto-rotational instability may occur after the formation of an accretion disk \citep{Tan04, Silk06}.

\section{Future probes with ALMA}
\begin{figure}
  \includegraphics[scale=0.45]{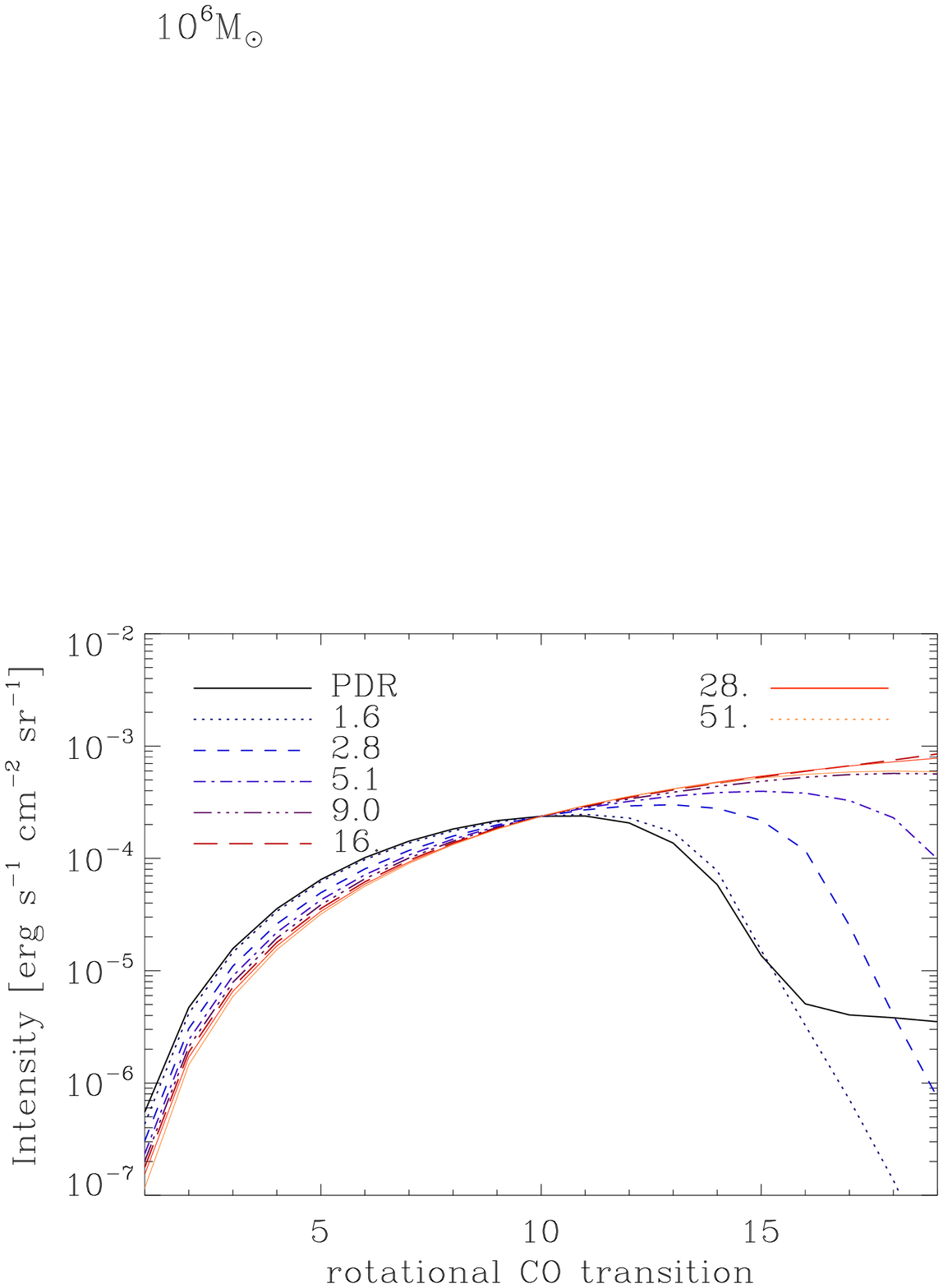}
  \caption{A comparison of the CO line SED in case of an intense starburst with $G_0=10^5$ with the corresponding SED for molecular clouds under X-ray irradiation, for different X-ray fluxes in erg~s$^{-1}$~cm$^{-2}$. The spectra are normalized such that they have the same intensity in the 10th transition.}
\label{fig:ALMA}
\end{figure}

As discussed by \citet{Spaans08} and \citet{SchleicherSpaans10}, the host galaxy of high-redshift black holes with $10^6$-$10^7$~M$_\odot$ may be probed with the upcoming mm/sub-mm telescope ALMA. At its largest baselines, it may resolve scales of $50-100$~pc at $z=8$, and thus probe the central regions of high-redshift quasars. Based on the CO line SED, it is possible to probe whether the chemistry there is driven by X-rays or the starburst (see Fig.~\ref{fig:ALMA}). It may further probe the dynamics with high-resolution line profiles.

\begin{theacknowledgments}
The research leading to these results has received funding from the European Community's Seventh Framework Programme (/FP7/2007-2013/) under grant agreement No 229517. Robi Banerjee is funded by the Emmy-Noether grant (DFG) BA 3607/1. {RSK thanks the German Science Foundation (DFG) for support via the Emmy Noether grant KL 1358/1.} DRGS and RSK also acknowledge subsidies from the DFG SFB 439 {\em Galaxies in the Early Universe}. DRGS, RSK, SS and thank for funding via {the Priority Programme 1177 {\em"Witnesses of Cosmic History:  Formation and evolution of black holes, galaxies and their environment"} of the German Science Foundation}. In addition, R.S.K.\   thanks for subsidies from the German {\em Bundesministerium f\"{u}r  Bildung und Forschung} via the ASTRONET project STAR FORMAT (grant  
05A09VHA) and from the {\em Landesstiftung Baden-W{\"u}rttemberg} via  their program International Collaboration II. RSK also thanks the KIPAC at Stanford University and the Department of Astronomy and Astrophysics 
at the University of California at Santa Cruz for their warm hospitality during a sabbatical stay in spring 2010. 

\end{theacknowledgments}





\begin{thebibliography}{26}
\expandafter\ifx\csname natexlab\endcsname\relax\def\natexlab#1{#1}\fi
\providecommand{\enquote}[1]{``#1''}
\expandafter\ifx\csname url\endcsname\relax
  \def\url#1{\texttt{#1}}\fi
\expandafter\ifx\csname urlprefix\endcsname\relax\def\urlprefix{URL }\fi
\providecommand{\eprint}[2][]{\url{#2}}

\bibitem[{Abel} et~al.(2002)]{Abel02}
T.~{Abel}, G.~L. {Bryan}, and M.~L. {Norman}, \emph{Science} \textbf{295},
  93--98 (2002).

\bibitem[{Bromm} and {Larson}(2004)]{Bromm04}
V.~{Bromm}, and R.~B. {Larson}, \emph{ARA\&A} \textbf{42}, 79--118 (2004),
  \eprint{arXiv:astro-ph/0311019}.

\bibitem[{Johnson} and {Bromm}(2007)]{Johnson07}
J.~L. {Johnson}, and V.~{Bromm}, \emph{MNRAS} \textbf{374}, 1557--1568 (2007),
  \eprint{arXiv:astro-ph/0605691}.

\bibitem[{Milosavljevic} et~al.(2008)]{Milosavljevic08}
M.~{Milosavljevic}, V.~{Bromm}, S.~M. {Couch}, and S.~P. {Oh}, \emph{ArXiv
  0809.2404}  (2008), \eprint{0809.2404}.

\bibitem[{Alvarez} et~al.(2009)]{Alvarez09}
M.~A. {Alvarez}, J.~H. {Wise}, and T.~{Abel}, \emph{ApJL} \textbf{701},
  L133--L137 (2009), \eprint{0811.0820}.

\bibitem[{Eisenstein} and {Loeb}(1995)]{Eisenstein95}
D.~J. {Eisenstein}, and A.~{Loeb}, \emph{ApJ} \textbf{443}, 11--17 (1995),
  \eprint{arXiv:astro-ph/9401016}.

\bibitem[{Bromm} and {Loeb}(2003)]{Bromm03}
V.~{Bromm}, and A.~{Loeb}, \emph{ApJ} \textbf{596}, 34--46 (2003),
  \eprint{arXiv:astro-ph/0212400}.

\bibitem[{Begelman} et~al.(2006)]{Begelman06}
M.~C. {Begelman}, M.~{Volonteri}, and M.~J. {Rees}, \emph{MNRAS} \textbf{370},
  289--298 (2006).

\bibitem[{Spaans} and {Silk}(2006)]{Spaans06}
M.~{Spaans}, and J.~{Silk}, \emph{ApJ} \textbf{652}, 902--906 (2006),
  \eprint{arXiv:astro-ph/0601714}.

\bibitem[{Li} et~al.(2005)]{LiMacLowKlessen05}
Y.~{Li}, M.-M. {Mac Low}, and R.~S. {Klessen}, \emph{ApJ} \textbf{626},
  823--843 (2005).

\bibitem[{Regan} and {Haehnelt}(2009)]{Regan09}
J.~A. {Regan}, and M.~G. {Haehnelt}, \emph{MNRAS} \textbf{396}, 343--353
  (2009), \eprint{0810.2802}.

\bibitem[{Shang} et~al.(2009)]{Shang09}
C.~{Shang}, G.~L. {Bryan}, and Z.~{Haiman}, \emph{MNRAS} pp. 1840--+ (2009),
  \eprint{0906.4773}.

\bibitem[{Omukai}(2001)]{Omukai01}
K.~{Omukai}, \emph{ApJ} \textbf{546}, 635--651 (2001),
  \eprint{arXiv:astro-ph/0011446}.

\bibitem[{Glover} and {Savin}(2009)]{Glover09}
S.~C.~O. {Glover}, and D.~W. {Savin}, \emph{MNRAS} \textbf{393}, 911--948
  (2009), \eprint{0809.0780}.

\bibitem[{Schleicher} et~al.(2010)]{Schleicher10b}
D.~R.~G. {Schleicher}, M.~{Spaans}, and S.~C.~O. {Glover}, \emph{ApJL}
  \textbf{712}, L69--L72 (2010), \eprint{1002.2850}.

\bibitem[{Dijkstra} et~al.(2008)]{Dijkstra08}
M.~{Dijkstra}, Z.~{Haiman}, A.~{Mesinger}, and S.~{Wyithe}, \emph{ArXiv
  e-prints}  (2008), \eprint{0810.0014}.

\bibitem[{Maki} and {Susa}(2004)]{Maki04}
H.~{Maki}, and H.~{Susa}, \emph{ApJ} \textbf{609}, 467--473 (2004),
  \eprint{arXiv:astro-ph/0403496}.

\bibitem[{Yoshida} et~al.(2008)]{Yoshida08}
N.~{Yoshida}, K.~{Omukai}, and L.~{Hernquist}, \emph{Science} \textbf{321},
  669-- (2008), \eprint{0807.4928}.

\bibitem[{Kazantsev}(1968)]{Kazantsev68}
A.~P. {Kazantsev}, \emph{Sov. Phys. JETP} \textbf{26}, 1031 (1968),
  \eprint{arXiv:astro-ph/0509149}.

\bibitem[{Brandenburg} and {Subramanian}(2005)]{Brandenburg05}
A.~{Brandenburg}, and K.~{Subramanian}, \emph{Phys. Rep.} \textbf{417}, 1--4
  (2005), \eprint{arXiv:astro-ph/0405052}.

\bibitem[{Arshakian} et~al.(2009)]{Arshakian09}
T.~G. {Arshakian}, R.~{Beck}, M.~{Krause}, and D.~{Sokoloff}, \emph{A\&A}
  \textbf{494}, 21--32 (2009), \eprint{0810.3114}.

\bibitem[{Schleicher et~al.}(2010)]{Schleicher10c}
D.~R.~G. {Schleicher et~al.}, \emph{ArXiv e-prints}  (2010),
  \eprint{1003.1135}.

\bibitem[{Tan} and {Blackman}(2004)]{Tan04}
J.~C. {Tan}, and E.~G. {Blackman}, \emph{ApJL} \textbf{603}, 401--413 (2004),
  \eprint{arXiv:astro-ph/0307455}.

\bibitem[{Silk} and {Langer}(2006)]{Silk06}
J.~{Silk}, and M.~{Langer}, \emph{MNRAS} \textbf{371}, 444--450 (2006),
  \eprint{arXiv:astro-ph/0606276}.

\bibitem[{Spaans} and {Meijerink}(2008)]{Spaans08}
M.~{Spaans}, and R.~{Meijerink}, \emph{ApJL} \textbf{678}, L5--L8 (2008),
  \eprint{arXiv:0803.2397}.

\bibitem[{Schleicher} et~al.(2010)]{SchleicherSpaans10}
D.~R.~G. {Schleicher}, M.~{Spaans}, and R.~S. {Klessen}, \emph{ArXiv e-prints
  1001.2118}  (2010), \eprint{1001.2118}.

\end{thebibliography}

\IfFileExists{\jobname.bbl}{}
 {\typeout{}
  \typeout{******************************************}
  \typeout{** Please run "bibtex \jobname" to optain}
  \typeout{** the bibliography and then re-run LaTeX}
  \typeout{** twice to fix the references!}
  \typeout{******************************************}
  \typeout{}
 }

\end{document}